\documentclass[aps,prd,superscriptaddress,notitlepage,nofootinbib,10pt]{revtex4-1}

\usepackage{graphicx}
\usepackage{amsmath}
\usepackage{amssymb}
\usepackage{xcolor}
\usepackage{bm}
\usepackage{hyperref}
\usepackage{slashed}


\begin{document}

\title{Second-order dissipative hydrodynamics for plasma with chiral asymmetry and vorticity}

\author{E.~V.~Gorbar}
%\email{gorbar@bitp.kiev.ua}
\affiliation{Department of Physics, Taras Shevchenko National Kiev University, Kiev, 03680, Ukraine}
\affiliation{Bogolyubov Institute for Theoretical Physics, Kiev, 03680, Ukraine}

\author{D.O. Rybalka}
%\email{drybalka@asu.edu}
\affiliation{Department of Physics, Arizona State University, Tempe, Arizona 85287, USA}

\author{I. A.~Shovkovy}
%\email{igor.shovkovy@asu.edu}
\affiliation{College of Integrative Sciences and Arts, Arizona State University, Mesa, Arizona 85212, USA}
\affiliation{Department of Physics, Arizona State University, Tempe, Arizona 85287, USA}

\date{\today}

\begin{abstract}
By making use of the chiral kinetic theory in the relaxation-time approximation, we derive an Israel-Stewart type 
formulation of the hydrodynamic equations for a chiral relativistic plasma made of neutral particles (e.g., neutrinos). 
The effects of chiral asymmetry are captured by including an additional continuity equation for the axial charge, 
as well as the leading-order quantum corrections due to the spin of particles. In a formulation of the chiral kinetic 
theory used, we introduce a symmetric form of the energy-momentum tensor that is suitable for the description 
of a weakly nonuniform chiral plasma. By construction, the energy and momentum are conserved to the same 
leading order in the Planck constant as the kinetic equation itself. By making use of such a chiral kinetic theory 
and the Chapman-Enskog approach, we obtain a set of second-order dissipative hydrodynamic equations. 
The effects of the fluid vorticity and velocity fluctuations on the dispersion relations of chiral vortical waves 
are analyzed.
\end{abstract}

\maketitle

\section{Introduction}

The concept of spin of elementary particles has been known for almost a century now. It has vast practical 
applications in science and technology. For massive particles, the spin is defined as the intrinsic angular 
momentum of the particle in its rest frame. For massless particles, such a frame is absent, and the spin can 
be defined indirectly via the particle's helicity as the spin projection on the particle's momentum. In the 
case of massless fermions, one can also use the concept of chirality instead of helicity. In fact, chirality 
and helicity  are same for particles (positive-energy states) and opposite of each other for antiparticles 
(negative-energy states). In a classical theory, the chirality of massless fermions is a conserved charge 
(quantum number). However, the chiral charge conservation is anomalous \cite{Adler,BellJackiw} and, 
thus, cannot be enforced in a quantum theory. 

In recent years, there was a surge of interest in chiral relativistic plasmas, in which chirality and/or chiral 
structure play a fundamental role. Theoretical studies of such plasmas revealed a number of unusual 
phenomena, including the chiral magnetic \cite{Fukushima:2008xe}, chiral separation \cite{Metlitski:2005pr} 
and chiral vortical \cite{Vilenkin:1979,Erdmenger:2008rm,Banerjee:2008th} effects among others. 
Their applications range from a possible generation of primordial magnetic fields in cosmology 
\cite{Boyarsky:2011uy,Tashiro:2012mf,Manuel:2015zpa,Hirono:2015rla,Gorbar:2016qfh,Gorbar:2016klv} 
to observable correlations of charged particles created in heavy-ion collisions
\cite{Kharzeev:2010gd,Burnier:2011bf,Gorbar:2011ya,Jiang:2015cva}, to unusual transport 
properties of Dirac/Weyl materials in condensed matter physics \cite{Li:2014bha}. For recent reviews, 
see Refs.~\cite{Kharzeev:2013ffa,Vafek:2013mpa,Miransky:2015ava,Burkov:2015,Landsteiner:2016led}.

In addition to the first-principles quantum-field theoretical methods in studies of chiral relativistic plasmas, 
several quasiclassical approaches were proposed as well. They include the chiral kinetic theory 
\cite{Son:2012zy,Stephanov:2012ki,Son:2012wh,Chen:2014cla,Manuel:2014dza} and chiral hydrodynamics 
\cite{Son:2009tf,Sadofyev:2010pr,Neiman:2010zi}. In particular, in the kinetic theory, which has an intermediate 
status between the microscopic approach and hydrodynamics, a chiral plasma is described in terms of a 
one-particle distribution function $f(x,p)$ in the phase space spanned by spatial coordinates and momenta. 
The fact that such a description may be possible for a plasma of massless fermions is interesting by itself. 
It is even more amazing, however, that the corresponding framework reproduces exactly the quantum chiral 
anomaly. 

The chiral anomaly is also taken into account in chiral hydrodynamics \cite{Son:2009tf,Sadofyev:2010pr,
Neiman:2010zi}, which describes local properties of plasma in terms of its conserved charge densities and 
the energy-momentum tensor. Among the three approaches discussed above, hydrodynamics is the least detailed 
one. Also, its range of validity is limited to the states of matter in the vicinity of equilibrium. Often, however, 
this is the most efficient and practical framework for the description of physical properties of matter in the 
long-wavelength limit. 

One of the most difficult tasks in a hydrodynamic description is the inclusion of dissipative effects. 
This is a particularly sensitive issue in the case of relativistic hydrodynamics, where a naive use of 
the gradient expansion is in conflict with the causality of the theory \cite{Lindblom,Denicol:2008ha,Pu:2009fj}.
The problem can be resolved by inclusion of higher moments of the distribution function \cite{Grad,Groot} 
beyond the basic hydrodynamic variables (e.g., the density of matter, energy density, and fluid velocity)
\cite{Muller:1967zza}. The second-order theory by Israel 
and Stewart \cite{Israel:1976tn,Israel:1979wp}, which employs additional purely damped degrees of 
freedom, solves the acausality problem and is widely used in the analysis of relativistic hydrodynamic 
systems. In principle, the dissipative form of hydrodynamics for chiral plasma can be derived from the 
chiral kinetic theory. Several generic algorithms for building such a theory are well known \cite{Kremer}. 
They utilize the definitions of conserved charges in the kinetic theory together with the gradient expansion 
for the distribution function in order to derive a consistent set of equations for hydrodynamic quantities. 
In order to obtain a closed set of equations, however, certain approximations are usually required. For example, 
one can use the moment expansion \cite{Denicol:2012} or the Chapman-Enskog method \cite{Chapman:book} 
and truncate the expansion at a given finite number of moments and gradients. 

In this paper, we will derive a closed set of dissipative hydrodynamic equations for a relativistic plasma by using 
the Chapman-Enskog method and following a truncation method similar to that in 
Refs.~\cite{Denicol:2010,Jaiswal:2013a,Jaiswal:2013,Jaiswal:2015mxa}, but paying a special attention to the effects due to 
chiral asymmetry and fluid vorticity.\footnote{A different approach to study the role of spin polarization in
relativistic plasmas was presented in Ref.~\cite{Montenegro:2017rbu}.} In this approach, instead 
of the 14-moment approximation of the original Israel-Stewart theory, an iterative solution of the Boltzmann 
equation is employed in order to derive the dissipative evolution equations. In particular, one of the novel 
features of our analysis will be the inclusion of the effects associated with a fluid vorticity in a chiral 
plasma.   

The paper is organized as follows. The key details of the chiral kinetic theory in the relaxation-time 
approximation are presented in Sec.~\ref{kinetic-theory}. The dissipative hydrodynamic equations are 
derived in Sec.~\ref{hydrodynamics}. In Sec.~\ref{sec:CVW}, we discuss several types of solutions 
in the form of attenuated propagating waves that involve the oscillations of chirality. The summary 
of the main results and general conclusions are given in Sec.~\ref{conclusion}. Some technical 
details and derivations are presented in Appendix~\ref{app:details}. In this paper we use units in 
which the speed of light is $c=1$.

\section{Chiral kinetic theory for plasma with non-uniform flow}
\label{kinetic-theory}

The starting point in our analysis is the chiral kinetic theory \cite{Son:2012zy,Stephanov:2012ki,Son:2012wh,
Chen:2014cla,Manuel:2014dza} in the relaxation-time approximation. The governing equations of such a theory are 
generalized Boltzmann equations for the distribution functions of chiral (Weyl) fermions. The corresponding 
particles can be of the left-hand ($\lambda=-1$) or right-hand chirality ($\lambda=+1$). Also, in view
of the relativistic nature of the system, the semi-classical framework at hand will not be complete without introducing 
both particles ($\chi=+1$) and antiparticles ($\chi=-1$) as independent species. In general, therefore, the chiral 
kinetic theory has four different distribution functions for the description of all four species of particles, 
$f_{\lambda,\chi}(x,p)$, where $\lambda=\pm1$ is the chirality and $\chi=\pm1$ is the sign of energy. 
In order to simplify the notation, we will suppress the indices $\lambda$ and $\chi$ in most 
formulas below.  

As already stated, we will use the chiral kinetic theory with a relaxation-time collision term as a starting 
point in the  derivation of the second-order dissipative chiral hydrodynamics. The relativistic form of the 
relaxation-time approximation was developed in Ref.~\cite{Anderson-Witting}. However,
it should be noted that the chiral kinetic theory in the relaxation-time approximation, while providing a great 
toy model, appears to be in conflict with the Lorentz covariance of the theory. In fact,
it is argued in Ref.~\cite{Chen:2015} that a collision term consistent with the Lorentz covariance should necessarily be nonlocal. 
For the purposes of this study, however, we will ignore this deficiency of the relaxation-time approximation 
in order to explore the structure of the theory in the simplest possible framework. 

For the purposes of this study, we require that the chiral kinetic equation be valid up to the linear 
order in $\hbar$ (or, equivalently, in spin). While such a form was proposed in Ref.~\cite{Chen:2015}, it
has to be recast in a format that allows one to describe a plasma with a spatially inhomogeneous flow 
velocity. By implementing the relaxation-time approximation as in Ref.~\cite{Anderson-Witting}, 
we write the kinetic equation in the following form:
\begin{equation}
\label{eq:kinetic_eq}
 p^\mu \partial_\mu f + (\partial_\mu S^{\mu\nu}) \partial_\nu f = -\frac{p \cdot u}{\tau}(f-f_{\rm eq}),
\end{equation}
where $p^\mu$ is the four-momentum of the particle, $u^\mu$ is the timelike four-velocity of the local plasma
flow (by assumption, $u^\mu u_\mu = 1$), and $\tau$ is the relaxation time. The spin tensor $S^{\mu\nu}$ 
\cite{Chen:2015} and the equilibrium distribution function $f_{\rm eq}(x,p)$ are defined as follows:
\begin{eqnarray} 
S^{\mu\nu} &=& \lambda\frac{\hbar}{2} \frac{\varepsilon^{\mu\nu\alpha\beta} p_\alpha u_\beta}{p \cdot u} 
\label{def-Smunu} ,\\
f_{\rm eq} &=& \frac{1}{1 + e^{\beta (\varepsilon_{p,{\rm eq}} - \chi \mu_\lambda)}} .
\label{def-f_eq} 
\end{eqnarray}
As is easy to check, the only nonvanishing components of the spin tensor in the local rest frame 
of the fluid are the spatial components: $S^{ij} = \lambda \hbar \varepsilon^{ijk} p^k/2|\mathbf{p}|$. 
Note that, in a general frame determined by the four-velocity $u^\mu$, the antisymmetric 
spin tensor satisfies the following relations: $u_\mu S^{\mu\nu} = p_\mu S^{\mu\nu} = 0$. 

The equilibrium distribution function (\ref{def-f_eq}) is defined in terms of the local values of the 
temperature $T\equiv 1/\beta$ and the chiral chemical potentials $\mu_\lambda$.
(Instead of using the chemical potentials $\mu_\lambda=\mu+\lambda\mu_5$, it may be also 
convenient to use the number-density and axial-charge density chemical potentials $\mu$ 
and $\mu_5$, respectively.) In this formalism, the dispersion relation 
for a chiral fermion is given by
\begin{equation}
 \varepsilon_{p,{\rm eq}} = \chi p^\mu u_\mu + \lambda\frac{\hbar}{2}\frac{p\cdot\omega}{p\cdot u}, 
\end{equation}
where the last term accounts for the spin contribution to the particle's energy connected with a nonzero 
vorticity of the flow $\omega^\mu \equiv \frac{1}{2} \varepsilon^{\mu\alpha\beta\gamma} u_\alpha \partial_\beta u_\gamma$. 
Here we use the conventional notation $\varepsilon^{\mu\nu\alpha\beta}$ for the four-dimensional 
Levi-Civita symbol. In the fluid rest frame, the vorticity takes its usual nonrelativistic form: 
$\boldsymbol{\omega} = \boldsymbol{\nabla} \times \mathbf{u}/2$.

In terms of the distribution function, the fermion-number and the axial-charge current densities are defined by the 
following expressions \cite{Chen:2015}:
\begin{eqnarray}
\label{eq:definition_j}
 j^\mu = \sum_{\lambda} \int \left( p^\mu f + S^{\mu\nu}\partial_\nu f \right),\\
\label{eq:definition_j5}
 j_5^\mu = \sum_{\lambda} \lambda \int \left( p^\mu f + S^{\mu\nu}\partial_\nu f \right), 
\end{eqnarray}
where we introduced the shorthand notation for the Lorentz-invariant momentum integration and the 
particle-antiparticle summation over $\chi$
\begin{equation}
\int F(p_0,\mathbf{p}) = \sum_\chi \int\frac{d^4p}{(2\pi)^3} 2 \delta(p^\mu p_\mu) \theta(\chi p^0) F(p_0,\mathbf{p})
= \sum_\chi \int\frac{d^3\mathbf{p}}{(2\pi)^3|\mathbf{p}|} F(\chi |\mathbf{p}|,\mathbf{p}) .
\end{equation}
It should be noted that, in addition to the usual orbital (or convective) part described by the first term 
in Eqs.~(\ref{eq:definition_j}) and (\ref{eq:definition_j5}), the definition of currents also contains a 
magnetization contribution connected with the spin. In the fluid rest frame, the latter for the 
fermion-number current takes the standard form of the curl of magnetization  \cite{Son:2012zy,Chen:2014cla}, 
i.e., $\boldsymbol{\nabla}\times \bm{\mathcal M}$, where 
$\bm{\mathcal M} \equiv \frac{\hbar}{2}  \sum_{\lambda} \lambda  \int \hat{\mathbf{p}} f$ and
$\hat{\mathbf{p}}\equiv \mathbf{p}/|\mathbf{p}|$. 

In this study, we neglect the correction to the current densities associated with the so-called side 
jumps during the collisions \cite{Chen:2015}. There are two reasons for this. First, it is not clear 
whether such a correction is meaningful and how to account for it in the kinetic equation when the 
relaxation-time approximation is used. Second, we assume that the relaxation time $\tau$ is 
rather large and, thus, the current corrections due to the side jumps are small. 

In terms of the particle distribution function, the energy-momentum tensor is defined as follows:
\begin{equation}
\label{eq:definition_T}
 T^{\mu\nu} = \sum_{\lambda}  \int
 \left( 
  p^\mu p^\nu f 
  + \frac{1}{2} p^\mu S^{\nu\alpha} \partial_\alpha f
  + \frac{1}{2} p^\nu S^{\mu\alpha} \partial_\alpha f
 \right).
\end{equation}
It is important to note that, unlike the case of the currents, there is no room for an ``axial" counterpart of 
the energy-momentum tensor. In the context of hydrodynamics, as we will see later, this is intimately 
connected with the fact that the two chiral components of the plasma should have the same temperature, 
even if they are characterized by different chemical potentials.

The definition in Eq.~(\ref{eq:definition_T}) is a straightforward generalization of the energy-momentum
tensor introduced in Ref.~\cite{Son:2012zy} 
to the case of a plasma with non-uniform flow. Just like the charge density, the energy-momentum tensor 
(\ref{eq:definition_T}) contains both orbital and spin contributions. By construction, this tensor is manifestly 
symmetric. Such a symmetric form of the tensor appears natural because the kinetic theory can be viewed as a 
semiclassical approximation to the microscopic quantum-field theoretical description, in which the corresponding 
tensor can be always symmetrized. A symmetric form of the energy-momentum tensor is also the most 
``physical'' from the viewpoint of general relativity; see Ref.~\cite{Hehl:1976vr}. It is interesting to mention, 
however, that hydrodynamics may allow for a nonzero antisymmetric part in the energy-momentum 
tensor (i.e., the torque tensor) which can be connected, for example, with the spin \cite{Halbwachs}. 
While we do not explore such a possibility here, it is intriguing to suggest that a spin-related torque 
tensor might be induced in a chiral plasma made of Weyl fermions.

\section{Hydrodynamic equations}
\label{hydrodynamics}

In this section, we derive a closed set of hydrodynamic equations for an inhomogeneous chiral 
plasma slightly out of local equilibrium. The corresponding local state is described in terms of the 
hydrodynamic variables $T$, $\mu_\lambda$, and $u^\mu$. Within the chiral kinetic theory, which
is the starting point in our derivation, these variables are sufficient to specify the local equilibrium 
distribution function (\ref{def-f_eq}).

In essence, the hydrodynamic equations are the continuity equations for conserved quantities, 
such as the energy and momentum, as well as various conserved charges. In order to close the 
corresponding system of equations, one should also add a number of constitutive relations. 
One of such relations is the equation of state that relates the energy density $\epsilon$ with 
the pressure $P$ of the fluid. In a relativistic plasma at hand, the latter is given by 
$P = \epsilon/3$. [Note that the corresponding equation of state also follows from 
the definition of the energy-momentum tensor in Eq.~(\ref{eq:definition_T}) combined with its 
representation in terms of the pressure and energy density; see Eq.~(\ref{eq:T_decomposition}) 
below.]

The hydrodynamic equations in a chiral plasma are governed by continuity equations for the current densities 
$j^\mu$ and $j_5^\mu$, as well as the energy-momentum tensor $T^{\mu\nu}$. Before deriving the 
equations for $j^\mu$, $j_5^\mu$, and $T^{\mu\nu}$, however, let us first discuss their vector/tensor structure. 
As usual, we will decompose these quantities using projections onto the Lorentz subspaces parallel 
and perpendicular to the four-velocity $u^\mu$. From a physics viewpoint, there are several possibilities 
for the choice of the four-velocity $u^\mu$ connected, e.g., with the energy flow (Landau frame)
or the particle flow (Eckart frame). While we will keep $u^\mu$ arbitrary for now, later we will see that 
the consistency of our hydrodynamic equations will single out a modified version of the Landau frame
\cite{Anderson-Witting}. The latter for particles with spin may differ from the usual Landau frame by corrections of order $\hbar$. 
In general, the current densities $j^\mu$ and $j_5^\mu$, and the energy-momentum tensor $T^{\mu\nu}$ 
have the following decompositions:
\begin{eqnarray}
 \label{eq:j_decomposition}
 j^\mu &=& n u^\mu + \nu^\mu, \\
 \label{eq:j5_decomposition}
 j_5^\mu &=& n_5 u^\mu + \nu_5^\mu, \\
 \label{eq:T_decomposition}
 T^{\mu\nu} &=& 
 \epsilon u^\mu u^\nu - \Delta^{\mu\nu} P + (h^\mu u^\nu + u^\mu h^\nu) + \pi^{\mu\nu}  ,
\end{eqnarray}
where $n \equiv u_\mu j^\mu$ and $n_5 \equiv u_\mu j_5^\mu$ are the fermion-number and axial-charge densities, 
$\nu^\mu \equiv \Delta^{\mu\nu} j_\nu$ and $\nu_5^\mu \equiv \Delta^{\mu\nu} j_{5,\nu}$ are the corresponding 
diffusion current densities perpendicular to the fluid four-velocity \cite{Denicol:2012,Jaiswal:2013,Jaiswal:2015mxa}, 
$\epsilon \equiv u_\mu u_\nu T^{\mu\nu}$ is the energy density, 
$P \equiv - \Delta_{\mu\nu} T^{\mu\nu}/3$ is the thermodynamic pressure, 
$h^\mu \equiv \Delta^\mu_\alpha u_\beta T^{\alpha\beta}$ is the momentum density
or the energy-flow density, and 
$\pi^{\mu\nu} \equiv \Delta^{\mu\nu}_{\alpha\beta} T^{\alpha\beta}$ is the shear stress tensor. 
By definition, the projector onto the subspace perpendicular to the fluid four-velocity is 
$\Delta^{\mu\nu} \equiv g^{\mu\nu} - u^\mu u^\nu$, 
and the traceless symmetric projector used in the definition of the shear stress tensor is 
$\Delta^{\mu\nu}_{\alpha\beta} \equiv
 (1/2) \Delta^\mu_\alpha \Delta^\nu_\beta 
+ (1/2) \Delta^\mu_\beta \Delta^\nu_\alpha
- (1/3) \Delta^{\mu\nu} \Delta_{\alpha\beta}$.

It should be noted that, the viscous pressure $\Pi$, which would shift the thermodynamic pressure $P\to P+\Pi$
in Eq.~(\ref{eq:T_decomposition}), is absent. From a physics viewpoint, such a correction would capture the 
effects of bulk viscosity. In the case of a nearly scale-invariant chiral plasma of (massless) fermions, however, 
the approximation with $\Pi= 0$ is well justified. Indeed, the bulk viscosity vanishes in scale-invariant theories.
Of course, in realistic models such as high-temperature QCD, the corresponding property is not exact because 
of subtle quantum effects. Nevertheless, as direct calculations in Ref.~\cite{Arnold:2006fz} show, the bulk viscosity 
is negligible compared to shear viscosity. Therefore, in the following we will completely ignore the effects due to the 
viscous pressure. 

In order to derive a closed system of hydrodynamic equations from the chiral kinetic theory, we use the 
approximation similar to that used in Refs.~\cite{Israel:1979wp,Denicol:2012,Jaiswal:2013,Jaiswal:2015mxa}. 
The relevant moments are $\epsilon$, $n$, $n_5$, $P$, $\nu^\mu$, $\nu^\mu_5$, $h^\mu$, and $\pi^{\mu\nu}$.
(Because of the spin contributions, strictly speaking, these quantities are not simple moments 
of the distribution function.) In the kinetic theory, the corresponding quantities can be expressed 
in terms of the particle distribution function, using the definitions for the current densities and the 
energy-momentum tensor in Eqs.~(\ref{eq:definition_j}), (\ref{eq:definition_j5}) and (\ref{eq:definition_T}). 

By making use of the general decomposition of the current densities (\ref{eq:j_decomposition}) 
and (\ref{eq:j5_decomposition}), and
the energy-momentum tensor (\ref{eq:T_decomposition}) in terms of independent moments, the 
continuity equations $\partial_\mu j^\mu=0$, $\partial_\mu j_5^\mu=0$, and $\partial_\nu T^{\mu\nu}=0$ 
take the following form:
\begin{eqnarray}
\label{eq:continuity_1}
 \dot{n} + n \partial_\mu u^\mu + \partial_\mu \nu^\mu &=& 0, \\
\label{eq:continuity5_1}
 \dot{n}_5 + n_5 \partial_\mu u^\mu + \partial_\mu \nu_5^\mu &=& 0, \\
\label{eq:continuity_2}
 \dot{\epsilon} + (\epsilon + P) \partial_\mu u^\mu + \partial_\mu h^\mu + u_\mu \dot{h}^\mu - \pi^{\mu\nu} \partial_\mu u_\nu &=& 0, \\
\label{eq:continuity_3}
 (\epsilon + P) \dot{u}^\alpha - \nabla^\alpha P + h^\mu \partial_\mu u^\alpha + h^\alpha (\partial_\mu u^\mu)
 + \Delta^\alpha_\mu \dot{h}^\mu 
  + \Delta^\alpha_\mu \partial_\nu \pi^{\mu\nu} &=& 0,
\end{eqnarray}
where we introduced the comoving and transverse derivatives as $\dot{A} \equiv u^\mu \partial_\mu A$ 
and $\nabla^\alpha P \equiv \Delta^{\alpha\mu} \partial_\mu P$, respectively. In dissipative regime, 
Eqs.~(\ref{eq:continuity_1})--(\ref{eq:continuity_3}) are not sufficient to describe unambiguously the 
hydrodynamic behavior of plasma. These equations should be supplemented by the equations for 
functions $\nu^\mu$, $\nu_5^\mu$, $h^\mu$, and $\pi^{\mu\nu}$. As is clear, their derivation 
would be impossible without additional information about the microscopic processes responsible for 
dissipative effects. In this study, the corresponding details will be supplied by the chiral kinetic theory 
in the relaxation-time approximation. 

Before attempting to derive the equations that govern the dynamics of dissipative functions, we
should first discuss the generic constraints that the chiral kinetic theory imposes on the hydrodynamic 
variables $T$, $\mu_\lambda$, and $u^\mu$. The corresponding variables determine the equilibrium 
distribution function, see Eq.~(\ref{def-f_eq}), and thus define the local state of equilibrium in plasma. 
It remains to determine, however, the connection between these variables and the out-of-equilibrium 
hydrodynamic functions that satisfy Eqs.~(\ref{eq:continuity_1})--(\ref{eq:continuity_3}). The needed 
relations can be established by analyzing the continuity equations, i.e., $\partial_\mu j^\mu=0$,
$\partial_\mu j_5^\mu=0$, and $\partial_\nu T^{\mu\nu}=0$, for $j^{\mu}$, $j_5^{\mu}$, and $T^{\mu\nu}$ 
given by Eqs.~(\ref{eq:definition_j}), (\ref{eq:definition_j5}), and (\ref{eq:definition_T}), respectively, in 
the framework of the chiral kinetic theory where the kinetic equation (\ref{eq:kinetic_eq}) plays the 
role of a microscopic equation.

Let us first discuss the implication of the continuity equations for the current densities. 
By making use of the definitions in Eqs.~(\ref{eq:definition_j}) and (\ref{eq:definition_j5}), 
it is straightforward to derive the following relation:
\begin{eqnarray}
\label{eq:continuity} 
\partial_\mu j^\mu &=& - \frac{1}{\tau} (n-n_{\rm eq}), \\
\label{eq:continuity5}
\partial_\mu j_5^\mu &=& - \frac{1}{\tau} (n_5-n_{5,{\rm eq}}), 
\end{eqnarray}
where we used the definitions for the densities $n\equiv \sum_{\lambda} \int (p\cdot u)f$ 
and $n_5\equiv \sum_{\lambda} \lambda \int (p\cdot u)f$. The explicit expressions for the
corresponding quantities in equilibrium are obtained by making use of the equilibrium 
distribution function in Eq.~(\ref{def-f_eq}), i.e.,
\begin{eqnarray}
n_{\rm eq} &\equiv& \sum_{\lambda} \left(\frac{\mu_\lambda^3}{6\pi^2} + \frac{\mu_\lambda T^2}{6}  \right)
=\frac{\mu\left(\mu^2+3\mu_5^2+\pi^2 T^2\right)}{3\pi^2} ,
\label{eq:n_equilibrium} \\
n_{5,{\rm eq}} &\equiv& \sum_{\lambda}\lambda \left(\frac{\mu_\lambda^3}{6\pi^2} + \frac{\mu_\lambda T^2}{6} \right)
=\frac{\mu_5\left(\mu_5^2+3\mu^2+\pi^2 T^2\right)}{3\pi^2} .
\label{eq:n5_equilibrium}
\end{eqnarray}
As follows from Eqs.~(\ref{eq:continuity}) and (\ref{eq:continuity5}), the requirements of the fermion-number 
and axial-charge conservation, $\partial_\mu j^\mu=0$ and $\partial_\mu j_5^\mu=0$, give
\begin{eqnarray}
n = n_{\rm eq}, & \quad\quad\quad\quad & n_5 = n_{5,{\rm eq}} .
\label{eq:constraint_1} 
\end{eqnarray}
These equations can be interpreted as definitions of the out-of-equilibrium charge densities in 
terms of given local values of the chemical potentials $\mu_\lambda$, or vice versa, as the 
equations that define $\mu_\lambda$ in terms of the local charge densities $n$ and $n_5$. 

By applying the same method to the definition in Eq.~(\ref{eq:definition_T}), we can also calculate
the divergence of the energy-momentum tensor in the chiral kinetic theory. The corresponding 
details are presented in Appendix~\ref{app:conservation}. The final result reads
\begin{equation}
\partial_\nu T^{\mu\nu} 
 = -\frac{u^\mu }{\tau} \left( \epsilon - \epsilon_{\rm eq}+\frac{\hbar}{2} \omega_\alpha
 (\nu_5^\alpha-\nu^\alpha_{5,{\rm eq}})\right) 
 -\frac{1}{\tau} \left(h^\mu -h^\mu_{\rm eq} 
  - \frac{\hbar}{4} \epsilon^{\mu\alpha\beta\gamma} u_\alpha \dot{u}_\beta 
  (\nu_{5,\gamma}-\nu_{5,{\rm eq},\gamma})\right) + O(\hbar^2),
\label{energy-momentum-conservation}
\end{equation}
where we used Eq.~(\ref{eq:constraint_1}). Because of the chiral vortical effect, the equilibrium axial-charge
current density is nonzero in the presence of a background vorticity, i.e., $\nu^\mu_{5,{\rm eq}} \propto 
\hbar\omega^\mu$. Considering that the corresponding result is already linear in $\hbar$, it contributes to the 
right-hand side of Eq.~(\ref{energy-momentum-conservation}) only at the quadratic order in $\hbar$. Therefore, 
while we will formally keep the equilibrium current density in similar expressions below, it would be consistent
to set $\nu^\mu_{5,{\rm eq}} \simeq 0$ there.

By separating the two independent projections 
with respect to the four-velocity $u^\mu$ and enforcing the continuity equation $\partial_\nu T^{\mu\nu}=0$, 
we then arrive at the following constraints:
\begin{eqnarray}
\label{eq:constraint_2}
\epsilon +\frac{\hbar}{2} \omega_\mu \nu_5^\mu
 &=& \epsilon_{\rm eq} + \frac{\hbar}{2} \omega_\mu \nu^\mu_{5,{\rm eq}} , \\
\label{eq:constraint_3}
h^\mu -  \frac{\hbar}{4} \epsilon^{\mu\alpha\beta\gamma} u_\alpha \dot{u}_\beta \nu_{5,\gamma}
 &=& h^\mu_{\rm eq} -
  \frac{\hbar}{4} \epsilon^{\mu\alpha\beta\gamma} u_\alpha \dot{u}_\beta \nu_{5,{\rm eq},\gamma}  ,
\end{eqnarray}
where, by definition, the equilibrium quantities are given by
\begin{eqnarray}
 \nu_{\rm eq}^\mu &\equiv& \hbar \omega^\mu\sum_{\lambda}\lambda \left(\frac{\mu_\lambda^2}{4\pi^2} + \frac{T^2}{12}\right)
 =\frac{\mu\mu_5}{\pi^2} \hbar \omega^\mu , 
\label{eq:equilibrium_current}\\
 \nu_\textrm{5,eq}^\mu &\equiv& \hbar \omega^\mu\sum_{\lambda} \left(\frac{\mu_\lambda^2}{4\pi^2} + \frac{T^2}{12}\right)
       = \frac{3(\mu^2+\mu_5^2)+\pi^2 T^2 }{6\pi^2} \hbar \omega^\mu, 
\label{eq:equilibrium_current5}\\
 \epsilon_{\rm eq} &\equiv & \sum_{\lambda} \left(\frac{\mu_\lambda^4}{8\pi^2} + \frac{\mu_\lambda^2 T^2}{4} + \frac{7\pi^2 T^4}{120}\right)
 =\frac{\mu^4+6\mu^2\mu_5^2+\mu_5^4}{4\pi^2}+\frac{T^2}{2}(\mu^2+\mu_5^2)+\frac{7\pi^2T^4}{60} , 
\label{eq:equilibrium_energy}\\
 h_{\rm eq}^\mu &\equiv& \hbar \omega^\mu n_{5,{\rm eq}}= \frac{\mu_5\left(\mu_5^2+3\mu^2+\pi^2 T^2\right)}{3\pi^2} \hbar \omega^\mu  .
\label{eq:equilibrium_momentum}
\end{eqnarray}
The constraint in Eq.~(\ref{eq:constraint_2}) for the energy density is analogous to the relations in Eq.~(\ref{eq:constraint_1}). 
This is most evident in the limit of vanishing vorticity or in the absence of axial current density, when the second term on 
each side of Eq.~(\ref{eq:constraint_2}) is trivial. Then, the equation $\epsilon=\epsilon_{\rm eq} $ allows one to 
define the out-of-equilibrium energy density in terms of a given local value of the temperature $T$ or, alternatively, 
to determine the value of $T$ in terms of the local energy density $\epsilon$. Because of the sum over $\lambda$, 
Eq.~(\ref{eq:constraint_2}) gives a single relation that allows one to determine only one (common) local 
temperature $T$ for both chiral components of the plasma. This is a natural consequence of the definition for the 
energy-momentum tensor in Eq.~(\ref{eq:definition_T}), which includes the contributions of both chiralities. 

Now let us turn to the second constraint, given by Eq.~(\ref{eq:constraint_3}). It can be viewed as the 
relation that defines the four-velocity $u^\mu$ of a generalized Landau frame, in which the hydrodynamic
equations derived from the kinetic theory are consistent with the momentum conservation. This 
agrees with a similar constraint (i.e., $h^\mu = 0$) obtained in a model without chiral asymmetry in 
Refs.~\cite{Anderson-Witting,Jaiswal:2013}. Indeed, after taking into account that all $\hbar$-terms 
in Eq.~(\ref{eq:constraint_3}) vanish at $n_5=0$ and $\nu^{\mu}_5=0$, we obtain the standard 
relation that defines the Landau frame: $h^\mu = 0$. Because of the $\hbar$ corrections, however, 
our definition of the generalized Landau frame in Eq.~(\ref{eq:constraint_3}) is different. It would be 
interesting to investigate whether the underlying reasons for the difference  is related to the subtleties 
of defining a thermodynamically preferred frame in Ref.~\cite{Becattini:2014yxa}.

It is interesting that constraints (\ref{eq:constraint_2}) and (\ref{eq:constraint_3}) contain corrections of order 
$\hbar$ when there is a nonzero background vorticity. In essence, the corresponding corrections to the energy 
and momentum densities are the consequences of chirality, which locks the orbital motion of particles with their 
spin. 

Even after taking into account Eqs.~(\ref{eq:constraint_1}), (\ref{eq:constraint_2}), 
and (\ref{eq:constraint_3}), we still need to determine the equations that govern the evolution of the dissipative 
functions $\nu^\mu$ and $\pi^{\mu\nu}$. Here we will follow the approach similar to that 
in Ref.~\cite{Jaiswal:2013} in order to derive the missing equations. We start by rewriting the kinetic 
equation in the following form:
\begin{eqnarray}
 \dot{f} + \frac{f}{\tau} = \frac{f_{\rm eq}}{\tau} - \frac{1}{p \cdot u} p^\rho \nabla_\rho f 
  - \frac{1}{p \cdot u} (\partial_\sigma S^{\sigma\rho}) \partial_\rho f.
\label{dissipation-effects}
\end{eqnarray}
Using the definitions for the dissipative functions, $\nu^\mu = \Delta^\mu_\nu j^\nu$, 
$\nu_5^\mu = \Delta^\mu_\nu j_5^\nu$, and $\pi^{\mu\nu} = \Delta^{\mu\nu}_{\alpha\beta} T^{\alpha\beta}$,
we can express their comoving derivatives in the following form:
\begin{eqnarray}
 \dot{\nu}^{\langle\mu\rangle} &=& 
 - \dot{u}^\mu n
 + \Delta^\mu_\nu \sum_{\lambda} \int \left( p^\nu + S^{\nu\alpha} \partial_\alpha \right) \dot{f}
 - \Delta^\mu_\nu \sum_{\lambda}\int S^{\nu\alpha} (\partial_\alpha u^\beta) \partial_\beta f 
 + \Delta^\mu_\nu \sum_{\lambda}\int \dot{S}^{\nu\alpha} \partial_\alpha f,
 \label{eq:dot_nu}
 \\
 \dot{\nu}_5^{\langle\mu\rangle} &=& 
 - \dot{u}^\mu n_5
 + \Delta^\mu_\nu \sum_{\lambda} \lambda \int \left( p^\nu + S^{\nu\alpha} \partial_\alpha \right) \dot{f}
 - \Delta^\mu_\nu \sum_{\lambda} \lambda \int S^{\nu\alpha} (\partial_\alpha u^\beta) \partial_\beta f 
 + \Delta^\mu_\nu \sum_{\lambda} \lambda \int \dot{S}^{\nu\alpha} \partial_\alpha f,
 \label{eq:dot_nu5}
 \\
 \label{eq:dot_pi}
 \dot{\pi}^{\langle\mu\nu\rangle} &=& - 2 \Delta^{\mu\nu}_{\alpha\beta} h^\alpha \dot{u}^\beta
 + \Delta^{\mu\nu}_{\alpha\beta} \sum_\lambda \int \left( p^\alpha p^\beta + p^{\alpha} S^{\beta\gamma} \partial_\gamma \right) \dot{f} 
 - \Delta^{\mu\nu}_{\alpha\beta} \sum_\lambda \int p^{\alpha} S^{\beta\gamma} (\partial_\gamma u^\delta) \partial_\delta f \nonumber\\
 &+& \Delta^{\mu\nu}_{\alpha\beta} \sum_\lambda \int p^{\alpha} \dot{S}^{\beta\gamma} \partial_\gamma f ,
\end{eqnarray}
where, by definition, the quantities with the Lorentz indices in angle brackets are the projections 
of the corresponding quantities onto the subspace orthogonal to the four-velocity, i.e., 
$\dot{\nu}^{\langle\mu\rangle} \equiv \Delta^\mu_\alpha \dot{\nu}^\alpha$ and
$\dot{\pi}^{\langle\mu\nu\rangle} \equiv \Delta^{\mu\nu}_{\alpha\beta} \dot{\pi}^{\alpha\beta}$. 
The use of projectors here is needed in order to force the dissipative current densities and the 
shear stress tensor to remain consistent with their generic definitions. This can be also viewed 
as a necessary condition for a self-consistent truncation of the evolution equations. 

By making use of the kinetic equation (\ref{dissipation-effects}), the relations for the comoving derivatives,
Eqs.~(\ref{eq:dot_nu})--(\ref{eq:dot_pi}), can be equivalently rewritten as follows:
\begin{eqnarray}
 \dot{\nu}^{\langle\mu\rangle} + \frac{\nu^\mu}{\tau} &=& - \dot{u}^\mu n
 + \sum_\lambda \int \left( \Delta^\mu_\nu p^\nu + S^{\mu\nu} \partial_\nu \right)
  \left( 
   \frac{f_{\rm eq}}{\tau} 
   - \frac{1}{p \cdot u} p^\rho \nabla_\rho f 
   - \frac{1}{p \cdot u} (\partial_\sigma S^{\sigma\rho}) \partial_\rho f
  \right) \nonumber \\
 &-& (\partial_\nu u^\rho) \sum_\lambda \int S^{\mu\nu} \partial_\rho f
 + \Delta^\mu_\rho \sum_\lambda \int \dot{S}^{\rho\nu} \partial_\nu f,
\label{eq:continuity_nuX}
\\
\dot{\nu}_5^{\langle\mu\rangle} + \frac{\nu_5^\mu}{\tau} &=& - \dot{u}^\mu n_5
 + \sum_\lambda \lambda \int \left( \Delta^\mu_\nu p^\nu + S^{\mu\nu} \partial_\nu \right)
  \left( 
   \frac{f_{\rm eq}}{\tau} 
   - \frac{1}{p \cdot u} p^\rho \nabla_\rho f 
   - \frac{1}{p \cdot u} (\partial_\sigma S^{\sigma\rho}) \partial_\rho f
  \right) \nonumber \\
 &-& (\partial_\nu u^\rho) \sum_\lambda \lambda \int S^{\mu\nu} \partial_\rho f
 + \Delta^\mu_\rho \sum_\lambda \lambda \int \dot{S}^{\rho\nu} \partial_\nu f,
\label{eq:continuity_nu5X}
\\
\label{eq:continuity_piX}
 \dot{\pi}^{\langle\mu\nu\rangle} + \frac{\pi^{\mu\nu}}{\tau} &=& 
 - 2 \Delta^{\mu\nu}_{\alpha\beta} h^\alpha \dot{u}^\beta
 + \Delta^{\mu\nu}_{\alpha\beta} \sum_\lambda \int \left( p^\alpha p^\beta + p^{\alpha} S^{\beta\gamma} \partial_\gamma \right)
  \left(
   \frac{f_{\rm eq}}{\tau}
   - \frac{1}{p \cdot u} p^\rho \nabla_\rho f 
   - \frac{1}{p \cdot u} (\partial_\sigma S^{\sigma\rho}) \partial_\rho f
  \right) \nonumber\\
 &-& (\partial_\gamma u^\rho) \sum_\lambda \Delta^{\mu\nu}_{\alpha\beta} \int p^{\alpha} S^{\beta\gamma} \partial_\rho f
 + \Delta^{\mu\nu}_{\alpha\beta} \sum_\lambda \int p^{\alpha} \dot{S}^{\beta\gamma} \partial_\gamma f.
\end{eqnarray}
These equations for dissipative functions contain the distribution function $f$. In order to obtain a closed set of equations, 
the right-hand sides of the equations above should be reexpressed in terms of the hydrodynamic variables and dissipative functions. 
To achieve this, we replace the distribution function with its iterative solution in the form
\begin{equation}
 f \simeq f_{\rm eq}
 - \frac{\tau}{p\cdot u} p\cdot \partial f_{\rm eq}
\label{eq:f_iterational}
\end{equation}
and further approximate the equilibrium distribution function by its expansion to the linear order in $\hbar$,
\begin{equation}
 f_{\rm eq} \simeq f_0
 + \lambda\frac{\hbar}{2} \frac{p^\mu \omega_\mu}{p \cdot u} f_0^{\prime} + \dots,
\label{eq:f_eq_iterational}
\end{equation}
where $f_0$ is the equilibrium function at a vanishing vorticity and $f_0^{\prime}\equiv \partial f_0/\partial\varepsilon_{p}$.
Now, by using the moments of the equilibrium distribution function from Appendix \ref{app:integrals}, we rewrite
the evolution equations for the dissipative functions in the following form:
\begin{eqnarray}
\label{eq:hydro_nu}
 \dot{\nu}^{\langle\mu\rangle} + \frac{\nu^\mu}{\tau} &=& - \dot{u}^\mu n + \sum_\lambda
 \Bigg[
\frac{1}{3} \nabla^\mu I_3
 + \lambda\frac{\hbar}{\tau} \omega^\mu I_2\nonumber \\
 &-& \frac{2\tau}{5} \nabla^\mu (\partial\cdot u) I_3
 - \frac{7\tau}{15} (\partial\cdot u) \nabla^\mu I_3
 + \tau \dot{u}^\mu (\partial\cdot u) I_3
 + \frac{4\tau}{5} \dot{u}^\rho (\partial_\rho u^\mu) I_3
 - \frac{\tau}{3} \nabla^\mu \dot{I}_3\nonumber \\
 &-& \frac{7\tau}{15} (\nabla_\rho u^\mu) \nabla^\rho I_3
 - \frac{2\tau}{5} (\nabla^\mu u_\rho) \dot{u}^\rho I_3
 - \frac{2\tau}{15} (\nabla^\mu u^\rho) \partial_\rho I_3
 - \frac{\tau}{5} \Delta^\mu_\nu (\partial\cdot\partial u^\nu) I_3
 + \frac{\tau}{5} \Delta^\mu_\nu \ddot{u}^\nu I_3\nonumber \\
 &-& \lambda\frac{14\hbar}{15} \omega^\mu (\partial\cdot u) I_2
 - \lambda\frac{14\hbar}{15} \omega^\nu (\partial_\nu u^\mu) I_2
 + \lambda\frac{\hbar}{15} \omega^\nu (\nabla^\mu u_\nu) I_2
 - \lambda\frac{2\hbar}{3} \omega^\mu \dot{I}_2\nonumber \\
 &+& \lambda\frac{\hbar}{6} \varepsilon^{\mu\nu\alpha\beta} u_\alpha \dot{u}_\beta \partial_\nu I_2
 - \lambda\frac{\hbar}{3} \varepsilon^{\mu\nu\alpha\beta} u_\beta (\partial_\nu u^\rho) (\partial_\rho u_\alpha) I_2\Bigg],
\end{eqnarray}
\begin{eqnarray}
\label{eq:hydro_nu5}
\dot{\nu}^{\langle\mu\rangle}_5 + \frac{\nu^\mu_5}{\tau} &=& - \dot{u}^\mu n_5 + \sum_\lambda
 \lambda\Bigg[
\frac{1}{3} \nabla^\mu I_3
 + \lambda\frac{\hbar}{\tau} \omega^\mu I_2\nonumber \\
 &-& \frac{2\tau}{5} \nabla^\mu (\partial\cdot u) I_3
 - \frac{7\tau}{15} (\partial\cdot u) \nabla^\mu I_3
 + \tau \dot{u}^\mu (\partial\cdot u) I_3
 + \frac{4\tau}{5} \dot{u}^\rho (\partial_\rho u^\mu) I_3
 - \frac{\tau}{3} \nabla^\mu \dot{I}_3\nonumber \\
 &-& \frac{7\tau}{15} (\nabla_\rho u^\mu) \nabla^\rho I_3
 - \frac{2\tau}{5} (\nabla^\mu u_\rho) \dot{u}^\rho I_3
 - \frac{2\tau}{15} (\nabla^\mu u^\rho) \partial_\rho I_3
 - \frac{\tau}{5} \Delta^\mu_\nu (\partial\cdot\partial u^\nu) I_3
 + \frac{\tau}{5} \Delta^\mu_\nu \ddot{u}^\nu I_3\nonumber \\
 &-& \lambda\frac{14\hbar}{15} \omega^\mu (\partial\cdot u) I_2
 - \lambda\frac{14\hbar}{15} \omega^\nu (\partial_\nu u^\mu) I_2
 + \lambda\frac{\hbar}{15} \omega^\nu (\nabla^\mu u_\nu) I_2
 - \lambda\frac{2\hbar}{3} \omega^\mu \dot{I}_2\nonumber \\
 &+& \lambda\frac{\hbar}{6} \varepsilon^{\mu\nu\alpha\beta} u_\alpha \dot{u}_\beta \partial_\nu I_2
 - \lambda\frac{\hbar}{3} \varepsilon^{\mu\nu\alpha\beta} u_\beta (\partial_\nu u^\rho) (\partial_\rho u_\alpha) I_2\Bigg],
\end{eqnarray}
\begin{eqnarray}
\label{eq:hydro_pi}
 \dot{\pi}^{\langle\mu\nu\rangle} + \frac{\pi^{\mu\nu}}{\tau} &=& 
 - 2 \Delta^{\mu\nu}_{\alpha\beta} h^\alpha \dot{u}^\beta
 + \Delta^{\mu\nu}_{\alpha\beta} \sum_\lambda
 \Bigg[
  \frac{8}{15} (\partial^\alpha u^\beta) I_4 \nonumber \\
  &-& \frac{32\tau}{35} (\partial^\alpha u^\beta) (\partial\cdot u) I_4
  - \frac{8\tau}{15} \partial^\alpha (\dot{u}^\beta I_4)
  - \frac{16\tau}{35} (\nabla_\rho u^\alpha) (\nabla^\rho u^\beta) I_4
  - \frac{8\tau}{21} (\partial^\alpha u^\rho) (\partial^\rho u^\beta) I_4 \nonumber \\
  &+& \frac{2\tau}{15} \partial^\alpha \partial^\beta I_4
  - \frac{2\tau}{3} (\partial^\alpha u^\beta) \dot{I}_4
  + \frac{8\tau}{105} (\partial^\alpha u_\rho) (\partial^\beta u^\rho) I_4 \nonumber \\
  &+& \lambda \frac{\hbar}{5} (\partial^\alpha \omega^\beta) I_3 
  + \lambda\frac{7\hbar}{15} \omega^\alpha \partial^\beta I_3 
  + \lambda\frac{\hbar}{5} \dot{u}^\alpha \omega^\beta I_3 
  + \lambda\frac{\hbar}{10} \varepsilon^{\beta\sigma\rho\delta} u_\delta \partial_\sigma (I_3 \nabla^\alpha u_\rho) \nonumber \\
  &+& \lambda\frac{\hbar}{10} \varepsilon^{\beta\sigma\rho\delta} u_\delta (\partial_\sigma u^\alpha) \partial_\rho I_3
  + \lambda\frac{\hbar}{5} \dot{u}^\alpha \varepsilon^{\beta\sigma\rho\delta} u_\rho (\partial_\sigma u_\delta) I_3
  + \lambda\frac{\hbar}{5} \varepsilon^{\beta\sigma\rho\delta} u_\sigma \dot{u}_\rho (\partial_\delta u^\alpha) I_3
 \Bigg].
\end{eqnarray}
As is easy to check, these equations for dissipative functions are finally sufficient to close the whole system of equations 
of the second-order dissipative hydrodynamics. Indeed, we have Eqs.~(\ref{eq:continuity_1})--(\ref{eq:continuity_3}) and 
(\ref{eq:constraint_3}) for hydrodynamic variables $n$, $n_5$, $\epsilon$, $u^{\mu}$, and $h^{\mu}$. Note also that 
the thermodynamic pressure is defined by the corresponding constitutive equation, $P = \epsilon/3$. The corresponding 
equations are supplemented by Eqs.~(\ref{eq:hydro_nu})--(\ref{eq:hydro_pi}) for functions $\nu^{\mu}$, $\nu^{\mu}_5$, and $\pi^{\mu\nu}$.
According to Eqs.~(\ref{eq:I2_calculated})--(\ref{eq:I4_calculated}) in Appendix \ref{app:integrals}, quantities $I_2$, $I_3$, 
and $I_4$ on the right-hand side of Eqs.~(\ref{eq:hydro_nu})--(\ref{eq:hydro_pi}) are expressed through the local equilibrium 
chemical potentials $\mu$, $\mu_5$ and temperature $T$, which in turn could be expressed through the local values of $n$, $n_5$, and 
$\epsilon$, respectively; see the constraints in Eqs.~(\ref{eq:constraint_1}) and (\ref{eq:constraint_2}).

The right-hand side of the equations for dissipative functions can be further simplified by making use of the following 
first-order relations:
\begin{eqnarray}
	\nu^\mu &=& \sum_\lambda \left[\lambda\hbar\omega^\mu I_2 + \frac{\tau}{3} \nabla^\mu I_3 - \tau \dot{u}^\mu I_3 \right]
	+ O(\partial^2) 
\label{nu-first-order}
,\\
	\nu^\mu_{5} &=& \sum_\lambda \lambda\left[\lambda\hbar\omega^\mu I_2 + \frac{\tau}{3} \nabla^\mu I_3 - \tau \dot{u}^\mu I_3 \right]
	+ O(\partial^2), 
\label{nu5-first-order} \\
	\pi^{\mu\nu} &=& \sum_\lambda \frac{8\tau}{15} \Delta^{\mu\nu}_{\alpha\beta} (\partial^\alpha u^\beta) I_4 + O(\partial^2),
\label{pi-first-order}
\end{eqnarray}
which follow from Eqs.~(\ref{eq:hydro_nu})--(\ref{eq:hydro_pi}). (Let us note in passing that the above 
first-order relations define the diffusion constant and the shear viscosity in terms of the relaxation time:
$D=\tau/3$ and $\zeta = 8\tau \epsilon/15$, respectively.) Indeed, by making use of these equations as well as the
continuity equations in the leading order in derivatives, we can reexpress most of the terms with an explicit dependence 
on the relaxation time in Eqs.~(\ref{eq:hydro_nu})--(\ref{eq:hydro_pi}) in terms of the hydrodynamic functions themselves. 
After doing this, the final set of equations for dissipative functions takes a simpler form, i.e.,
\begin{eqnarray}
\label{eq:hydro_nu-final-1}
 \dot{\nu}^{\langle\mu\rangle} + \frac{\nu^\mu- \nu^\mu_{\rm eq}}{\tau} &=& - \dot{u}^\mu n + \frac{1}{3} \nabla^\mu n   
 - \frac{n}{\epsilon+P} \Delta^{\mu\nu} \partial^\rho \pi_{\rho\nu}
	- \nu_\rho \omega^{\rho\mu}
	- (\partial\cdot u) \nu^\mu
	- \frac{9}{5} (\partial^{\langle\mu} u^{\rho\rangle})\nu_\rho
	+ \frac{14}{15} (\nabla^{\langle\mu} u^{\rho\rangle}) \nu_{{\rm eq},\rho}  \nonumber \\
	&-&\frac{2}{9} (\partial\cdot u) \nu^\mu_{\rm eq}
	-  \frac{2\hbar}{3} \omega^\mu \sum_\lambda \lambda \dot{I}_2
	+ \sum_\lambda \lambda \frac{\hbar}{6} \varepsilon^{\mu\nu\alpha\beta} \left[ u_\alpha \dot{u}_\beta \partial_\nu I_2 
	- 2 u_\beta (\partial_\nu u^\rho) (\partial_\rho u_\alpha) I_2 \right],
\end{eqnarray}
\begin{eqnarray}
\label{eq:hydro_nu5-final-1}
\dot{\nu}^{\langle\mu\rangle}_5 + \frac{\nu^\mu_5-\nu^\mu_{5,{\rm eq}}}{\tau} &=& - \dot{u}^\mu n_5 + \frac{1}{3} \nabla^\mu n_5  
 - \frac{n_5}{\epsilon+P} \Delta^{\mu\nu} \partial^\rho \pi_{\rho\nu}
	- \nu_{5,\rho} \omega^{\rho\mu}
	- (\partial\cdot u) \nu_5^\mu
	- \frac{9}{5} (\partial^{\langle\mu} u^{\rho\rangle}) \nu_{5,\rho} 
	+\frac{14}{15} (\nabla^{\langle\mu} u^{\rho\rangle}) \nu_{5,{\rm eq},\rho} \nonumber \\
	&-&\frac{2}{9} (\partial\cdot u) \nu^\mu_{5,{\rm eq}}
	-  \frac{2\hbar}{3} \omega^\mu \sum_\lambda \dot{I}_2
	+ \sum_\lambda  \frac{\hbar}{6} \varepsilon^{\mu\nu\alpha\beta} \left[ u_\alpha \dot{u}_\beta \partial_\nu I_2 
	- 2 u_\beta (\partial_\nu u^\rho) (\partial_\rho u_\alpha) I_2 \right],
\end{eqnarray}
\begin{eqnarray}
\label{eq:hydro_pi-final-1}
 \dot{\pi}^{\langle\mu\nu\rangle} + \frac{\pi^{\mu\nu}}{\tau} &=& - 2 h^{\langle\mu} \dot{u}^{\nu\rangle}
	+ 2 \pi_\rho^{\langle\mu} \omega^{\nu\rangle\rho}
	- \frac{10}{7} \pi_\rho^{\langle\mu} \sigma^{\nu\rangle\rho}
	- \frac{4}{3} \pi^{\mu\nu} \partial_\alpha u^\alpha
	+ \frac{8}{15} (\partial^{\langle\mu} u^{\nu\rangle}) \epsilon \nonumber \\
		&+& \frac{\hbar}{5} \left( (\partial^{\langle\mu} \omega^{\nu\rangle}) n_5 
		+ \frac{7}{3} \omega^{\langle\mu} \partial^{\nu\rangle} n_5 
		- \dot{u}^{\langle\mu} \omega^{\nu\rangle} n_5 \right)
		\nonumber \\
		&+&\frac{\hbar}{5} \Delta^{\mu\nu}_{\alpha\beta}\varepsilon^{\beta\sigma\rho\delta} \Bigg[  
		\frac{1}{2}  u_\delta \partial_\sigma (n_5 \nabla^\alpha u_\rho) 
		+ \frac{1}{2} u_\delta (\partial_\sigma u^\alpha) \partial_\rho n_5 
		+ u_\sigma \dot{u}_\rho (\partial_\delta u^\alpha) n_5
	\Bigg]  ,
\end{eqnarray}
where  $\sigma^{\mu\nu}=\partial^{\langle\alpha} u^{\beta\rangle}=\Delta^{\mu\nu}_{\alpha\beta} (\partial^\alpha u^\beta)$, 
$\omega^{\mu\nu}=(\nabla^\mu u^\nu - \nabla^\nu u^\mu)/2$, and 
$A^{\langle\mu\nu\rangle} \equiv \Delta^{\mu\nu}_{\alpha\beta} A^{\alpha\beta}$. 
In the derivation, we used the following relation:
\begin{equation} 
	\frac{1}{\epsilon + P} \Delta^{\mu\nu} \partial^\rho \pi_{\rho\nu} 
	= \frac{\tau}{5} \Delta^{\mu\nu} (\partial^2 u_\nu)
	- \frac{\tau}{5} \Delta^{\mu\nu} \ddot{u}_\nu
	+ \frac{\tau}{15} \nabla^\mu (\partial\cdot u)
	- \frac{3\tau}{5} \dot{u}^\mu (\partial\cdot u)
	+ \frac{3\tau}{5} \dot{u}^\rho (\partial_\rho u^\mu)
	+ \frac{4\tau}{5} \dot{u}^\rho (\nabla^\mu u_\rho) + O(\partial^3),
\end{equation}
which follows from Eq.~(\ref{pi-first-order}) as well as the first-order continuity equations. 

The set of second-order equations (\ref{eq:hydro_nu-final-1}), (\ref{eq:hydro_nu5-final-1}), and (\ref{eq:hydro_pi-final-1}) 
for dissipative functions in a chiral plasma is our main result. This is a generalization of the previous results of 
Refs.~\cite{Anderson-Witting,Denicol:2010,Jaiswal:2013a,Jaiswal:2013,Jaiswal:2015mxa}, 
which were obtained for massless plasmas without a chiral asymmetry (i.e., $n_5=0$ and $\nu_5^\mu=0$)
and without $\hbar$ corrections due to the spin. In the current study, in contrast, we treated the 
fermion chiralities as two components of a relativistic fluid. The (approximate) conservation of the axial charge in the 
chiral plasma gives rise to an additional continuity equation; see Eq.~(\ref{eq:hydro_nu5-final-1}). Moreover, the quantum 
effects of the chiral plasma are captured by the linear in $\hbar$ corrections in the second-order theory.

\section{Chiral vortical and sound waves}
\label{sec:CVW}

In order to illustrate how the hydrodynamic equations derived in the previous section could be used in practice, 
we discuss in this section its simplest solutions describing attenuated chiral vortical and sound waves that 
involve the oscillations of chirality. As we will see, a proper 
account of the fluid flow affects the properties of such a wave.

We begin our analysis by recalling that the existence of the chiral vortical wave is a direct consequence of 
the chiral vortical effect. In essence, the later states that a nonzero fluid vorticity in a chiral plasma induces 
the following fermion-number and axial-charge currents \cite{Landsteiner:2011iq,Landsteiner:2013aba,Gao:2012,
Golkar:2012kb}:
\begin{eqnarray}
\mathbf{j}&=&\frac{\mu\mu_5}{\pi^2}\,\bm{\omega},
\label{electric-current-CVE}
\\
\mathbf{j}_5&=&  \left(\frac{T^2}{6}+\frac{\mu^2+\mu^2_5}{2\pi^2}\right)\,\bm{\omega}.
\label{chiral-current-CVE}
\end{eqnarray}
Note that while the fermion-number current (\ref{electric-current-CVE}) exists only when both $\mu$ 
and $\mu_5$ are nonzero, the chiral current (\ref{chiral-current-CVE}) exists even if $\mu=\mu_5=0$ due to 
the $T^2$-term. The latter is related to the gravitational anomaly \cite{Landsteiner:2011iq,Landsteiner:2013aba,
Golkar:2015oxw}.
As suggested in Ref.~\cite{Jiang:2015cva}, an interplay between the fermion-number  and axial-charge 
fluctuations induced by the chiral vortical effect results in a gapless collective excitation that was called 
the chiral vortical wave. As we will see below, the inclusion of the hydrodynamic flow profoundly modifies 
this simple picture. 

To start with, let us note that the dissipative equations derived in the previous section reproduce the correct equilibrium 
expressions for the fermion-number and axial-charge currents given by Eqs.~(\ref{electric-current-CVE}) and 
(\ref{chiral-current-CVE}). Indeed, as the system approaches equilibrium, all gradient terms in Eqs.~(\ref{eq:hydro_nu-final-1}) 
and (\ref{eq:hydro_nu5-final-1}) vanish and their solutions take a particularly simple form: $\nu^{\mu} = \nu_{\rm eq}^{\mu} $ 
and $\nu_{5}^{\mu} = \nu_{5,{\rm eq}}^{\mu} $, where  the equilibrium currents are defined by 
Eqs.~(\ref{eq:equilibrium_current}) and (\ref{eq:equilibrium_current5}). As is easy to check, the latter 
coincide with the results in Eqs.~(\ref{electric-current-CVE}) and (\ref{chiral-current-CVE}) in the rest frame of 
the fluid $u^{\mu}=(1,0,0,0)$. In this connection, it should be noted that the currents $\nu^{\mu}$ and $\nu^{\mu}_5$ 
are generically defined as dissipative quantities. At the same time, the equilibrium chiral vortical effect currents $\nu_{\rm eq}^{\mu}$ 
and $\nu_{5,{\rm eq}}^{\mu}$ are nondissipative parts of $j^{\mu}$ and $j^{\mu}_5$. For simplicity of notations, 
however, we still include them in the dissipative functions $\nu^{\mu}$ and $\nu^{\mu}_5$.

In order to analyze the chiral vortical wave by using the hydrodynamic equations obtained in the previous section,
we choose the local background velocity of the fluid in the following form: 
\begin{equation}
u^\mu = u^\mu_0 + \epsilon^{\mu\nu\alpha\beta} x_\nu u_{0\alpha} \bar{\omega}_\beta,
\label{four-velocity}
\end{equation}
where the first term describes a uniform motion and the second one describes a rotation. We will 
assume that relation (\ref{four-velocity}) is valid for a sufficiently slow rotation and sufficiently small distances 
$L \ll |\bar{\omega}|^{-1}$. The above expression for the four-velocity is normalized in the usual way, 
$u^\mu u_\mu = 1$.
Up to quadratic terms in vorticity, which are negligible in the case of a slow rotation, 
the normalization condition for $u^\mu$ is valid if the four-vectors $u^\mu_0$ and $\bar{\omega}^\mu$ 
satisfy $u_0^\mu\bar{\omega}_{\mu} = 0$ and $u_0^\mu u_{0,\mu} = 1$. One should also note 
that, to leading order, the four-vector $\bar{\omega}^\mu$ coincides with the definition of vorticity given by 
$\omega^\mu \equiv \frac{1}{2} \varepsilon^{\mu\alpha\beta\gamma} u_\alpha \partial_\beta u_\gamma$. 

Let us search for a solution to hydrodynamic equations in the form of a propagating wave. In the most general 
case, the chemical potentials, temperature, and fluid velocity will oscillate around their average values, 
i.e.,
\begin{equation}
 \delta \mu(x) = e^{-ikx} \delta \mu_0, \qquad
 \delta \mu_5(x) = e^{-ikx} \delta \mu_{5,0}, \qquad
 \delta T(x) = e^{-ikx} \delta T_0, \qquad
 \delta u^\mu(x) = e^{-ikx} \delta u_0^\mu,
\end{equation}
where $k^\mu$ is the wave vector, and $\delta \mu_0$, $\delta \mu_{5,0}$, $\delta T_0$, and $\delta u^\mu_0$ 
are the amplitudes of oscillations of the corresponding quantities. The requirement of normalization constrains 
the oscillations of the fluid velocity to be orthogonal to the background velocity, $u_\mu \delta u^\mu = 0$. 
This is automatically satisfied for the waves with the fluid velocity oscillations along the direction of the vorticity, 
i.e., $\delta u^\mu \parallel \bar{\omega}^\mu$.

For the sake of simplicity, let us analyze the dissipative equations in the first-order theory. 
In this case, we find from Eqs.~(\ref{eq:hydro_nu-final-1})--(\ref{eq:hydro_pi-final-1}) that
\begin{eqnarray}
 \nu^\mu &=& \nu^\mu_{\rm eq} - \tau n \dot{u}^\mu + \frac{\tau}{3} \nabla^\mu n, \\
 \nu_5^\mu &=& \nu^\mu_{5,{\rm eq}} - \tau n_5 \dot{u}^\mu + \frac{\tau}{3} \nabla^\mu n_5, \\
 \pi^{\mu\nu} &=&  \frac{8\tau}{15} \epsilon   (\partial^{\langle\mu} u^{\nu\rangle}),
\end{eqnarray}
where we used the constraints (\ref{eq:constraint_1}), (\ref{eq:constraint_2}), and (\ref{eq:constraint_3}).
By substituting these expressions into the continuity equations (\ref{eq:continuity_1})--(\ref{eq:continuity_3}) 
and linearizing them in fluctuations, we derive the following system of coupled equations:
\begin{eqnarray}
 \sum_{z_i}\left( 
  \Omega \frac{\partial n_{\rm eq}}{\partial z_i } - i \frac{\tau}{3} k_\perp^2 \frac{\partial n_{\rm eq}}{\partial z_i } + k_\mu 
  \frac{\partial \nu ^\mu_{\rm eq}}{\partial z_i } 
 \right) \delta z_i 
 + n_{\rm eq} \left( 1 + i\tau\Omega \right) (k\cdot\delta u) - 2 \Omega  (\nu _{\rm eq}\cdot\delta u) &=& 0, 
 \label{CVW-eq1}
 \\
  \sum_{z_i}\left( 
  \Omega \frac{\partial n_{5,{\rm eq}}}{\partial z_i } - i \frac{\tau}{3} k_\perp^2 \frac{\partial n_{5,{\rm eq}}}{\partial z_i } + k_\mu 
  \frac{\partial \nu ^\mu_{5,{\rm eq}}}{\partial z_i } 
 \right) \delta z_i
 + n_{5,{\rm eq}} \left( 1 + i\tau\Omega \right) (k\cdot\delta u) - 2 \Omega  (\nu _{5,{\rm eq}}\cdot\delta u) &=& 0, 
 \label{CVW-eq2}
 \\
  \sum_{z_i}\left( 
  \Omega \frac{\partial \epsilon_{\rm eq}}{\partial z_i } + k_\mu \frac{\partial h ^\mu_{\rm eq}}{\partial z_i } 
 \right) \delta z_i
 + \frac{4}{3} \epsilon_{\rm eq} (k\cdot\delta u) - 3 \Omega  (h _{\rm eq}\cdot\delta u) &=& 0, 
 \label{CVW-eq3}
 \\
  \sum_{z_i}\left( 
  \Omega  \frac{\partial h ^\mu_{\rm eq}}{\partial z_i } - k_\perp^\mu \frac{1}{3} \frac{\partial \epsilon_{\rm eq}}{\partial z_i } 
 \right) \delta z_i
 + \left(
  \frac{4}{3} \epsilon_{\rm eq} \Omega  + \frac{3}{4} (k\cdot h _{\rm eq}) 
 \right) \delta u^\mu + h ^\mu_{\rm eq} (k\cdot\delta u) 
&& \nonumber\\
 + \frac{1}{4} (h _{\rm eq}\cdot\delta u) k_\perp^\mu  
 - \frac{i}{2} \hbar n_{5,{\rm eq}} \Omega  \varepsilon^{\mu\nu\alpha\beta} u_\nu k_\alpha \delta u_\beta 
 - i\tau \frac{8}{15} \Delta^{\mu\nu}_{\alpha\beta} k_\nu 
\epsilon_{\rm eq} k^\alpha \delta u^\beta
&=& 0 .
 \label{CVW-eq4}
\end{eqnarray}
Here we introduced the shorthand notations $\Omega=(k\cdot u)$ and 
$k_\perp^\mu=k^\mu - u^\mu (k\cdot u)$, and used the summation index 
$z_i=(\mu,\mu_5,T)$. The analysis of these equations is simplified in the rest frame 
with $u_0^\mu=(1,0,0,0)$ and $\bar{\omega}^\mu=(0,0,0,\bar{\omega})$. 
For a wave propagating along the direction of vorticity, the wave vector takes the 
form $k^\mu = \Omega u_0^\mu + k_z \hat{\omega}^\mu = (\Omega,0,0,k_z)$. 

The obtained system of homogeneous linear equations has nontrivial solutions only 
when the determinant of the corresponding matrix of coefficients vanishes. Thus, by solving
the characteristic equation, we obtain dispersion relations for four different types 
of waves: two sound waves and two modes that resemble chiral vortical waves. As
we will see, the latter differ from the simplified solutions of the chiral vortical waves 
\cite{Jiang:2015cva} because their propagation is profoundly affected by the 
hydrodynamic flow of the fluid itself. 

To the linear order in $\omega$ and $\tau$, the resulting dispersion relations for the sound waves are given by
\begin{equation}
\label{Omega-sound-wave}
 \Omega = \pm \frac{k_z}{\sqrt{3}} + \frac{3}{8} \hbar \bar{\omega} \frac{n_{5,{\rm eq}}}{\epsilon_{\rm eq}} k_z  + \frac{2}{15} i\tau k_z^2,
 %\Omega = \pm k_z  \left(\frac{1}{\sqrt{3}} \pm \frac{3 \hbar \bar{\omega} n_{5,{\rm eq}}}{8\epsilon_{\rm eq}} \right) - \frac{2}{15} i\tau k_z^2,
\end{equation}
where the second term is a vorticity correction to the usual speed of sound and the third term describes the  
attenuation of the sound wave. The dispersion relations of the chiral vortical waves read
\begin{equation}
 \Omega = \hbar\bar{\omega} v_1 k_z - \frac{1}{3} i\tau k_z^2, \qquad
 \Omega = \hbar\bar{\omega} v_2 k_z - \frac{1}{3} i\tau k_z^2,
\end{equation}
where $v_{1,2}$ are the roots of a quadratic equation $av^2+bv+c=0$ with the following coefficients:
\begin{eqnarray}
 a &=& \epsilon \Big[ 45 \left(\mu^2-\mu_5^2\right)^2 \left(\mu^2+\mu_5^2\right)+7 \pi ^6 T^6+27 \pi ^4 T^4 \left(\mu^2+\mu_5^2\right)+3 \pi ^2 T^2 \left(11 \mu^4+18 \mu^2 \mu_5^2+11 \mu_5^4\right) \Big],
 \\
 b &=& \frac{\mu_5}{10\pi^2} \Big[ 225 (\mu^2-\mu_5^2)\left(2 \mu^6+5 \mu^4 \mu_5^2+8 \mu^2 \mu_5^4+\mu_5^6\right)-14 \pi ^8 T^8-\pi ^6 T^6 \left(78 \mu^2+127 \mu_5^2\right) \nonumber\\
 &&-45 \pi ^4 \mu_5^2 T^4 \left(11 \mu^2+9 \mu_5^2\right)+15 \pi ^2 T^2 \left(30 \mu^6+5 \mu^4 \mu_5^2-72 \mu^2 \mu_5^4-43 \mu_5^6\right)\Big],
 \\
 c &=& \frac{3}{20\pi^2} \Big[ 75 \mu_5^8-4 \left(\pi ^3 \mu T^3+5 \pi  \mu^3 T\right)^2+225 \mu_5^6 \left(3 \mu^2+\pi ^2 T^2\right)-3 \mu_5^2 \left(5 \mu^2+\pi ^2 T^2\right)^2 \left(5 \mu^2+7 \pi ^2 T^2\right) \nonumber\\
 &&+5 \mu_5^4 \left(-75 \mu^4+13 \pi ^4 T^4+30 \pi ^2 \mu^2 T^2\right)\Big].
\end{eqnarray}
It is worth noting that there are two different modes of the chiral vortical wave. This result seems to qualitatively agree
with the dispersion relations obtained in Refs.~\cite{Abbasi:2015saa,Abbasi:2016rds}. From a physics point of view, 
they correspond to two opposite directions of propagation with respect to the vorticity. In general, the speeds 
of such waves are different. It is interesting to note that the corresponding waves have nonzero velocities $\hbar\bar{\omega}v_{1,2}$ 
even at $\mu=0$, which appears to contradict the prediction of Ref.~\cite{Kalaydzhyan:2016dyr}, where similar 
waves were analyzed. We may suggest that this is the result of using a more general scheme in this study, in 
which both the fermion-number and axial-charge conservations are enforced (see also Refs.~\cite{Abbasi:2015saa,Abbasi:2016rds}).

It is instructive to consider a special case of a plasma with the vanishing axial-charge chemical potential $\mu_5=0$. 
In this case, the dispersion relation for the sound waves is similar to that in Eq.~(\ref{Omega-sound-wave}), but has 
no correction due to vorticity. This should not be surprising for a plasma without a chiral asymmetry. As for the 
dispersion relations of the chiral vortical waves, they are given by the following explicit expression:
\begin{equation}
 \Omega = \pm \frac{\bar{\omega}\hbar T\mu(\pi^2T^2+5\mu^2)k_z}{2\pi\sqrt{5\epsilon_{\rm eq} (5\epsilon_{\rm eq}-2T^2\mu^2)(\pi^2T^2+3\mu^2)}} - \frac{1}{3} i\tau k_z^2.
\end{equation}
As is easy to check from Eqs.~(\ref{CVW-eq1})--(\ref{CVW-eq4}), the propagation of the chiral vortical waves is 
characterized by oscillations of all thermodynamic parameters. In fact, in this case, the explicit relations between 
their oscillation amplitudes are given by 
\begin{eqnarray}
 \delta \mu_{5,0} &=& \pm \frac{4\pi}{(\partial\epsilon_{\rm eq}/\partial T)} \sqrt{\frac{\epsilon_{\rm eq} (5\epsilon_{\rm eq}-2T^2\mu^2)}{5(\pi^2T^2+3\mu^2)}} \delta\mu_{0}, \\
 \delta T_{0} &=& - \frac{3n_{\rm eq}}{(\partial\epsilon_{\rm eq}/\partial T)} \delta\mu_{0}, \\
 \delta u_{0}^\mu &=& \mp \frac{\hbar\bar{\omega}^\mu}{\pi (\partial\epsilon_{\rm eq}/\partial T)} \sqrt{\frac{(5\epsilon_{\rm eq}-2T^2\mu^2)(\pi^2T^2+3\mu^2)}{5\epsilon_{\rm eq}}} \delta\mu_{0}.
\end{eqnarray}
As we see from the last equation, the chiral vortical wave is accompanied by oscillations of the fluid velocity 
along the direction of the vorticity. This is in addition to the usual oscillations of the fermion-number and axial-charge 
densities. With the model assumptions used here, we also see that the chiral vortical waves come with local 
oscillations of the temperature. At the same time, as is easy to check, the chiral vortical  waves do not drive 
the oscillations of the local energy density, $\delta \epsilon = 0$.

\section{Conclusion}
\label{conclusion}

In this study, we derived a closed system of second-order dissipative hydrodynamic equations that governs 
the evolution of a chiral plasma made of neutral particles such as neutrinos. The corresponding results can 
be applied for studies of the early states of protoneutron star evolution, where neutrinos are trapped in 
dense matter and achieve a hydrodynamic regime. In such a plasma, the effects of chirality could play 
an important role in driving an inverse cascade that may be relevant for the origin of the supernova 
explosion \cite{Yamamoto:2016zut}. The system of hydrodynamic equations obtained here can be used 
to numerically simulate the corresponding dynamics. While the corresponding detailed study is beyond the 
scope of this paper, we see that the presence of a chiral asymmetry modifies the hydrodynamic 
equations. It also appears that the leading-order quantum corrections due to spin of chiral particles can 
play profound effects in hydrodynamics, especially when combined with a chiral asymmetry and vorticity
of the fluid. 

By making use of the hydrodynamic equations, in this paper we also briefly addressed the modification of 
the chiral vortical waves associated with the fluid flow. In this part, for simplicity, we used the first-order 
theory. Of course, such an approximation is sufficient for the problem of the propagating modes with 
long wavelengths when there is no issue with the stability of solutions. We found that the 
propagation of the chiral vortical wave also induces oscillations of the local fluid velocity. As a result,
its dispersion relation differs from that predicted in a simplified model where only the oscillations of fermion-number 
and axial-charge densities are taken into account. Interestingly, we find that the local energy density does not 
oscillate during the propagation of the chiral vortical wave.

While the effects of electromagnetism were neglected in this study, there is no conceptual limitation
to take them into account. In fact, a generalization of the second-order hydrodynamic equations to 
the case of a chiral plasma made of charged particles is of great interest. The corresponding plasmas
play a profound role in cosmology, heavy-ion collisions, and even Dirac/Weyl materials. Obviously, 
many interesting phenomena may be expected from a nontrivial interplay of electromagnetic
fields and vorticity. In principle, the derivation of the corresponding equations is a straightforward although 
tedious task that we plan to address in the future. 

Another interesting extension of the current study would be the derivation of the third-order dissipative 
hydrodynamics \cite{Jaiswal:2013} and, perhaps, even the inclusion of the quantum corrections
beyond the leading order in the Planck constant. One should keep in mind, however, that 
additional corrections of quantum origin could be expected even in the chirally symmetric matter at the 
second order in $\hbar$ \cite{Becattini:2015nva}. Concerning the quantum corrections, it should be 
noted that, in the framework proposed in this study, the conservations of the energy and momentum 
were enforced only to the linear order in $\hbar$, or equivalently in spin. Of course, before attempting 
the inclusion of higher-order quantum corrections, this limitation should be lifted first. One of the promising 
approaches that could help to advance the problem is based on the use of the Wigner function 
\cite{Gao:2012,Chen:2013}. 

In view of the obvious importance of preserving the Lorentz covariance in relativistic 
models, it will be desirable to generalize the current analysis to the case of models with nonlocal 
collision integrals consistent with the Lorentz covariance. While the problem is expected to be 
much more challenging technically, it may not be hopeless \cite{Hidaka:2016yjf}. 

\begin{acknowledgments}
The authors would like to thank P.~Sukhachov and G.~Torrieri for useful comments and discussions.
The work of E.V.G. was supported in part by the Swiss National Science Foundation, Grant No. SCOPE IZ7370-152581,
and by the Program of Fundamental Research of the Physics and Astronomy Division of the National Academy of Sciences of Ukraine.
The work of D.O.R. and I.A.S. was supported by the U.S. National Science Foundation under Grant No.~PHY-1404232.
\end{acknowledgments}

\appendix

\section{The energy-momentum conservation and other technical details}
\label{app:details}

In this appendix, we present selected technical details that support the results in the main 
text of the paper. 

\subsection{The energy-momentum conservation}
\label{app:conservation}

The conservation of the energy and momentum to the leading order in the Planck constant plays a
very important role in our analysis. So, here we present the derivation of the general relation for 
the energy-momentum tensor, i.e., $\partial_{\nu} T^{\mu\nu}=0$, in the chiral kinetic theory. As 
we show, this condition is not automatically satisfied in the relaxation-time approximation. It can 
be enforced by choosing a special frame of reference.  

From the definition of the energy-momentum tensor in Eq.~(\ref{eq:definition_T}), we obtain
\begin{eqnarray}
\partial_\nu T^{\mu\nu} 
 &=& \frac{1}{2} \int \left( \delta^\mu_\nu \delta^\kappa_\lambda - \delta^\mu_\lambda \delta^\kappa_\nu \right) 
  p^\lambda (\partial_\kappa S^{\nu\alpha}) \partial_\alpha f 
 - \frac{1}{\tau} \int p^\mu (p\cdot u) (f-f_{\rm eq}) 
 + \frac{1}{2} \int S^{\mu\alpha} \partial_\alpha (p\cdot\partial f) \nonumber\\
 &\simeq & \frac{1}{4} \int \varepsilon^{\mu\kappa\sigma\rho} \varepsilon_{\nu\lambda\sigma\rho}
  p^\lambda (\partial_\kappa S^{\nu\alpha}) \partial_\alpha f 
 - \frac{1}{\tau} \int p^\mu (p\cdot u) (f-f_{\rm eq}) 
 - \frac{1}{2\tau} \int S^{\mu\alpha} \partial_\alpha\left[(p\cdot u)(f-f_{\rm eq})\right],
\label{energy-momentum-appendix1}
\end{eqnarray}
where we temporarily omitted the sum over $\lambda$ and dropped the terms of order $\hbar^2$. 
(Note that $S^{\mu\alpha}$ is linear in $\hbar$.)
It is convenient to analyze the first integral on the right-hand side separately. By making use of the 
explicit form of the spin tensor $S^{\nu\alpha}$, we can rewrite the corresponding integrand as follows:
\begin{equation}
\frac{1}{4} \int \varepsilon^{\mu\kappa\sigma\rho} \varepsilon_{\nu\lambda\sigma\rho} p^\lambda (\partial_\kappa S^{\nu\alpha})
\partial_\alpha f 
  =\lambda\frac{\hbar}{8} \int \varepsilon^{\mu\kappa\sigma\rho} \delta^{\alpha\beta\gamma}_{\lambda\sigma\rho} 
  p^\lambda p_\beta \partial_\kappa \left( \frac{u_\gamma}{p \cdot u} \right) \partial_\alpha f,
\end{equation}
where $\delta^{\alpha\beta\gamma}_{\lambda\sigma\rho} = \epsilon^{\alpha\beta\gamma\delta}
\epsilon_{\lambda\sigma\rho\delta}$ is the generalized Kronecker symbol. Because of the Kronecker 
symbol, the contraction over index $\lambda$ 
will lead to one of the following three possibilities: (i) $p^\lambda \to p^\alpha$, (ii)  $p^\lambda \to p^\beta$, or
(iii)  $p^\lambda \to p^\gamma$. The latter two cases give vanishing results either because $p^\beta p_\beta =0$ 
(massless particles), or because $\partial_\kappa \left( \frac{p^\gamma u_\gamma}{p \cdot u} \right) = 0$. The only 
nontrivial contribution comes from the contraction that turns $p^\lambda$ into $p^\alpha$, i.e.,
\begin{equation}
\lambda\frac{\hbar}{4} \int \varepsilon^{\mu\kappa\beta\gamma}
  p^\alpha p_\beta \partial_\kappa \left( \frac{u_\gamma}{p \cdot u} \right) \partial_\alpha f
  \simeq -\frac{1}{2\tau} \int  \left(\partial_\kappa S^{\mu\kappa} \right) 
  (p \cdot u)(f-f_{\rm eq}),
\label{energy-momentum-appendix3}
\end{equation}
where we used the kinetic equation and dropped the terms of order $\hbar^2$ on the right-hand side.
Now, by combining the results in Eqs.~(\ref{energy-momentum-appendix1}) and (\ref{energy-momentum-appendix3}),
we finally obtain
\begin{eqnarray}
\partial_\nu T^{\mu\nu} 
 &=& - \frac{1}{\tau} \int p^\mu (p\cdot u) (f-f_{\rm eq}) 
 - \lambda\frac{\hbar}{4\tau} \int  \varepsilon^{\mu\alpha\nu\beta} p_\nu \partial_\alpha \left[ u_\beta 
 (f-f_{\rm eq})\right] + O(\hbar^2) 
 \nonumber\\
 &=& -\frac{1}{\tau} \left[ u^\mu (\epsilon - \epsilon_{\rm eq}) +(h^\mu -h^\mu_{\rm eq}) 
  - \lambda \frac{\hbar}{2} \omega^{\mu} (n  - n_{\rm eq})
  - \lambda \frac{\hbar}{4} \epsilon^{\mu\alpha\beta\gamma} (\partial_\alpha u_\beta) (\nu_\gamma - \nu_\gamma^{\rm eq})
 \right] + O(\hbar^2).
\label{energy-momentum-appendix4}
\end{eqnarray}
Here we used the chiral kinetic theory definitions for the energy density $\epsilon \equiv u_\mu u_\nu T^{\mu\nu}$, 
momentum density $h^\mu\equiv \Delta^\mu_\alpha u_\beta T^{\alpha\beta}$, 
charge density $n \equiv u_\mu j^\mu$, and 
current density $\nu^\mu \equiv \Delta^{\mu\nu} j_\nu$ that follow directly from the definitions for 
the number-density current (\ref{eq:definition_j}) and the energy-momentum tensor (\ref{eq:definition_T}).

After restoring the sum over $\lambda$, the result in Eq.~(\ref{energy-momentum-appendix4}) 
can be rewritten in the following equivalent form:
\begin{equation}
\partial_\nu T^{\mu\nu} 
 = -\frac{u^\mu }{\tau} \left( \epsilon - \epsilon_{\rm eq}+\frac{\hbar}{2} \omega_\mu 
  (\nu_5^\mu-\nu^\mu_{5,{\rm eq}})\right) -\frac{1}{\tau} \left( h^\mu -h^\mu_{\rm eq} 
  - \frac{\hbar}{4} \epsilon^{\mu\alpha\beta\gamma} u_\alpha \dot{u}_\beta (\nu_{5,\gamma} -\nu_{5,{\rm eq},\gamma})\right) + O(\hbar^2),
\label{energy-momentum-appendix5}
\end{equation}
where we used the constraint in Eq.~(\ref{eq:constraint_1}) and separated the components along the 
four-vector $u^{\mu}$ from the projection perpendicular to $u^{\mu}$. Note the last term in the perpendicular 
component can be given in several alternative forms
\begin{equation}
\frac{\hbar}{4} \epsilon^{\mu\alpha\beta\gamma} u_\alpha \dot{u}_\beta (\nu_{5,\gamma} -\nu_{5,{\rm eq},\gamma}) 
=\Delta^\mu_\nu \frac{\hbar}{4} \epsilon^{\nu\alpha\beta\gamma} 
(\partial_\alpha u_\beta) (\nu_{5,\gamma} -\nu_{5,{\rm eq},\gamma}) 
= \frac{\hbar}{4} \epsilon^{\mu\alpha\beta\gamma} (\partial_\alpha u_\beta) (\nu_{5,\gamma} -\nu_{5,{\rm eq},\gamma})
 +\frac{\hbar}{2} u^\mu \omega^\gamma (\nu_{5,\gamma} -\nu_{5,{\rm eq},\gamma}).
\end{equation}

\subsection{Useful integrals}
\label{app:integrals}

In the calculation of moments of the distribution function, the following integrals are useful:
\begin{eqnarray}
 \int (p\cdot u)^n f_0 &=& - \sum_{\chi=\pm1} \frac{\Gamma(n+2)}{2\pi^2} \chi^n T^{n+2} \textrm{Li}_{n+2}
 \left( -e^\frac{\chi\mu_\lambda}{T} \right) \equiv I_{n+2},  \\
 \int (p\cdot u)^n p^\alpha f_0 &=& u^\alpha I_{n+3},  \\
 \int (p\cdot u)^n p^\alpha p^\beta f_0 &=& \left( -\frac{1}{3} g^{\alpha\beta} + \frac{4}{3} u^\alpha u^\beta \right) I_{n+4},  \\
 \int (p\cdot u)^n p^\alpha p^\beta p^\gamma f_0 &=& \left( - g^{(\alpha\beta} u^{\gamma)} + 2 u^\alpha u^\beta u^\gamma \right) I_{n+5},\\
 \int (p\cdot u)^n p^\alpha p^\beta p^\gamma p^\delta f_0 &=& 
  \left( \frac{1}{5} g^{(\alpha\beta} g^{\gamma\delta)} - \frac{12}{5} g^{(\alpha\beta} u^{\gamma} u^{\delta)}
  + \frac{16}{5} u^\alpha u^\beta u^\gamma u^\delta \right) I_{n+6},  \\
 \int (p\cdot u)^n p^{\mu_1} ... p^{\mu_5} f_0 &=& 
  \left( 
   g^{(\mu_1} g u^{\mu_5)} 
   - \frac{16}{3} g^{(\mu_1} u .. u^{\mu_5)} 
   + \frac{16}{3} u^{\mu_1} .. u^{\mu_5} 
  \right) I_{n+7},  \\
 \int (p\cdot u)^n p^{\mu_1} ... p^{\mu_6} f_0 &=& 
  \left(
   - \frac{1}{7} g^{(\mu_1} g g^{\mu_6)} 
   + \frac{24}{7} g^{(\mu_1} g u u^{\mu_6)} 
   - \frac{80}{7} g^{(\mu_1} u .. u^{\mu_6)}
   + \frac{64}{7} u^{\mu_1} .. u^{\mu_6}
  \right) I_{n+8},
\end{eqnarray}
where $f_0$ is the equilibrium function at a vanishing vorticity and the round brackets 
denote a symmetrization over all possible permutations, e.g., 
$A^{(\alpha}B^{\beta}C^{\gamma)}\equiv  
       ( A^{\alpha}B^{\beta}C^{\gamma} 
      + A^{\alpha}B^{\gamma}C^{\beta} 
      + A^{\beta}B^{\alpha}C^{\gamma} 
      + A^{\beta}B^{\gamma}C^{\alpha} 
      + A^{\gamma}B^{\beta}C^{\alpha} 
      + A^{\gamma}B^{\alpha}C^{\beta}
)/3!$.

It is easy to check that lower moments can be obtained from the higher ones multiplying the latter 
by the four-velocity $u^{\mu}$. Similar integral chains can be obtained also for derivatives of the 
distribution function $f^{\prime}_0=\partial f_0/\partial\varepsilon_{p}$ if one makes a substitution 
$I_n \rightarrow -(n-1)I_{n-1}$. For $f^{\prime\prime}_0$, the substitution is $I_n \rightarrow (n-1)(n-2)I_{n-2}$ 
and so on. As is easy to check, the explicit results for several lowest-order moments read
\begin{eqnarray}
 I_1 &=& \frac{\mu_\lambda}{2\pi^2},\label{eq:I1_calculated} \\
 I_2 &=& \frac{\mu_\lambda^2}{4\pi^2} + \frac{T^2}{12}, \label{eq:I2_calculated}\\
 I_3 &=& \frac{\mu_\lambda^3}{6\pi^2} + \frac{\mu_\lambda T^2}{6},  \label{eq:I3_calculated}\\
 I_4 &=& \frac{\mu_\lambda^4}{8\pi^2} + \frac{\mu_\lambda^2 T^2}{4} + \frac{7\pi^2 T^4}{120}.
\label{eq:I4_calculated}
\end{eqnarray}
Note that these moments satisfy the following recurrent relation: $\partial I_{n+1}/\partial\mu_\lambda=n I_n $.

\end{document}